\documentclass[prb,twocolumn]{revtex4-1}

\usepackage{amssymb}
\usepackage{amsmath}
\usepackage{graphicx}

\begin{document}

\title{Tailoring the magnetic properties of Fe asymmetric nanodots}

\author{B. Leighton$^{1}$, N. M. Vargas$^{1}$, D. Altbir$^{1,2}$, and J. Escrig$^{1,2}$}
\address{$^1$ Departamento de F\'{\i}sica, Universidad de Santiago de Chile, USACH,
Av. Ecuador 3493, Santiago, Chile} 
\address{$^2$ Center for the Development of Nanoscience and
Nanotechnology, CEDENNA, Av. Ecuador 3493, Santiago, Chile}

\begin{abstract}
Asymmetric dots as a function of their geometry have been investigated using three-dimensional (3D) object oriented micromagnetic framework (OOMMF) code. The effect of shape asymmetry of the disk
on coercivity and remanence is studied. Angular dependence of the remanence and coercivity is also addressed.
Asymmetric dots are found to reverse their magnetization by nucleation and
propagation of a vortex, when the field is applied parallel to the direction of asymmetry. However, complex reversal modes appear when the angle at which
the external field is applied is varied, leading to a non monotonic behavior
of the coercivity and remanence.
\end{abstract}

\maketitle

\section{Introduction}

During the last decade a great deal of attention has been focused on the
study of regular arrays of magnetic particles produced by nano-imprint
lithography. Besides the basic scientific interest in the magnetic
properties of these systems, lithographed magnetic nanostructures are good
candidates for the production of new magnetic devices, magnetic sensors and
logical devices\cite{CW00} as high density storage media\cite{CAK+98}. The
properties exhibited by these nanostructures are strongly dependent on the
geometry, and therefore great control of the shape is fundamental for the
understanding and applications of such materials\cite{Ross01}.

In the case of nanodots, two main reversal mechanisms for the magnetization
have been observed, vortex nucleation and coherent rotation \cite{RLS+09}.
Vortex states are characterized by in-plane and out-of-plane magnetization.
In-plane magnetization forms the vortex chirality, a clockwise or
counterclockwise magnetization rotation around a core. Out-of-plane
magnetization defines the polarity, the up or down direction of the vortex
core magnetization. In this way, vortices exhibit four different magnetic
states and can then store the information of four magnetic bits. Switching
of the vortex polarity by the application of small magnetic fields along the
dot axis suggests the possibility of using the vortex for high density
storage media. Methods to control chirality in single FM layer elements
exploit an asymmetry in the applied field, such as using a magnetic force
microscope tip \cite{MGF+07, JYP+10}, magnetic pulses \cite{GSM08}, or a
magnetic field gradient \cite{KYK+08}, as well as the magnetization history 
\cite{KRL+01}. Also the geometry can be used to control polarity and/or
chirality, and dots with slight geometric asymmetry \cite{SHZ01, WHW+08} or triangular nanodots \cite{JYP+10, VBM+06} have been used for this purpose.

On the other hand, some groups adopted the idea of asymmetric disks to
achieve control over the vortex chirality with an in-plane magnetic field 
\cite{SHZ01, WHW+08, VBM+06, TOA+05, GPB+07, DGL+10}. Following this idea,
Wu \textit{et al}. \cite{WHW+08} studied vortex nucleation, annihilation and
field distribution switching in 40-nm-thick Ni$_{80}$Fe$_{20}$ disk arrays,
with a diameter of 300 nm and different degrees of asymmetry. Their
measurements and micromagnetic simulations showed that the nucleation and
annihilation of vortices have a linear relation with respect to the ratio of the long/short asymmetry axes, while the switching field
distribution oscillates. More recently, Dumas \textit{et al}. \cite{DGL+10}
reported the synthesis and magnetic characterization of polycrystalline arrays of asymmetric
Co dots where the circular shape of all the dots had been broken in the same
way. In these arrays they showed how the vortices can be manipulated to
annihilate at particular sites under certain field orientations and cycling
sequences.

On another side, several works on magnetic nanoparticles focus on
coercivity, which is strongly dependent on geometric parameters. Although
current experimental facilities allow the fabrication of dots with a variety
of geometries, precision is still a problem. Therefore, the possibility of
controlling the coercivity by other means is highly desirable. Since
coercivity is directly related to the reversal mechanism, one alternative is
to induce different reversal modes by modifying external parameters, such as
the direction of the applied field.

Following these ideas, in this paper, micromagnetic simulations have been used
to investigate the angular dependence of the hysteresis and reversal modes
for non-interacting asymmetric dots as a function of their geometry. We focus on the
behavior of the coercive field and remanence and conclude that magnetic
fields applied along different directions on asymmetric dots are a possible
way of tailoring magnetic properties of nanodots.

\section{Micromagnetic background}

The theory of micromagnetism was developed by Brown Jr. about 50 years ago 
\cite{Brown63}. According to this model, a ferromagnetic system consisting
of a large number of individual magnetic spins is described using continuous
functions for the magnetization, the fields and the energies. Moreover, the
amplitude of the magnetization vector $\mathbf{M}(\mathbf{r})$ has to be
constant, but its orientation may change from one position to another. In
this approach, for a given magnetization distribution $\mathbf{M}(\mathbf{r}%
) $ the Gibbs free energy is

\begin{multline}
G\left( \mathbf{M}\right) =\int dV \\
\left[ A\left( \mathbf{\nabla m}\right) ^{2}-\frac{1}{2}\mu _{0}M_{s}\left( 
\mathbf{m}\cdot \mathbf{H}_{dem}\right) -\mu _{0}M_{s}\left( \mathbf{m}\cdot 
\mathbf{H}\right) \right] 
\end{multline}%
where $\mathbf{H}$ is an applied field, $A$ is the exchange constant and $%
\mathbf{H}_{dem}$ is the demagnetizing field. Using the variational principle to minimize the Gibbs free
energy with respect to the magnetization $\mathbf{M}$, the equation of the
stable equilibrium state is $\mathbf{m}\times \mathbf{H}_{eff}=0$, where the
effective field $\mathbf{H}_{eff}$ is defined as $\mathbf{H}_{eff}=-dG/d%
\mathbf{M}$. Because the magnetostatic interaction is of long range,
analytical solutions of micromagnetic problems can only be obtained for
samples of simple shape and making use of simplifying assumptions \cite%
{Aharoni96}. Because of their symmetry, ring geometries are particularly
suited for such analytical calculations \cite{BLS+06, LEA+06}. However, for
systems having a complex geometry, like asymmetric rings, numerical
simulations are required.

The magnetization dynamics is governed by the Landau-Lifshitz-Gilbert
equation (LLG) \cite{Gilbert55}

\begin{equation}
\frac{d\mathbf{M}}{dt}=-\gamma \mathbf{M\times H}_{eff}+\frac{\alpha }{M_{s}}%
\mathbf{M}\times \frac{d\mathbf{M}}{dt}
\end{equation}%
where $\gamma $ is the gyromagnetic ratio of the free electron spin and $%
\alpha $ is a phenomenological damping constant. The equation describes a
combined precession and relaxation motion of the magnetization in an
effective field $\mathbf{H}_{eff}$. The calculations presented here have
been computed by Oriented Micro Magnetic Framework software (OOMMF)\cite%
{Oommf}. Thus, the ferromagnetic system is spatially divided into cubic
cells and within each cell the magnetization is assumed to be uniform. In
order to assure a good description of the magnetization details, the size of
the mesh has to be smaller than the exchange length of the material, defined
by $l_{ex}=\sqrt{2A/\mu _{0}M_{s}^{2}}$. The choice of the discretization
scheme is validated by the fact that the numerical roughness (generated by
the square mesh representation) corresponds to the real imperfections on the
lateral ring surface, arising for example from the resolution of the
patterning methods used.

\section{Sample specification}

Our starting point is a uniform circular dot with diameter $d$ and height $h$. We introduce asymmetries in these dots by cutting specific sections characterized by the parameter $\delta=r/R$, as illustrated in Fig.
1. A symmetric dot is characterized by $\delta =1.0$, while a semicircular
one is represented by $\delta =$ 0.0. To simulate the magnetic properties we used micromagnetic simulations, assuming that the dot-dot distance is large enough to consider every dot as independent\cite{MAL+10}. For our simulations we use the typical Fe parameters: saturation magnetization $M_{s}=1.859 \times 10^6$ A/m, exchange stiffness constant $A=45.78 \times 10^{-12}$ J/m, and a mesh size of 2 nm, where spins are free to rotate in three
dimensions. Since we are interested in polycrystalline samples, anisotropy is very small and can be safely neglected%
\cite{MAL+10}. In all the cases the damping parameter was chosen as 0.5. 

\begin{figure}[h]
\begin{center}
\includegraphics[width=7cm]{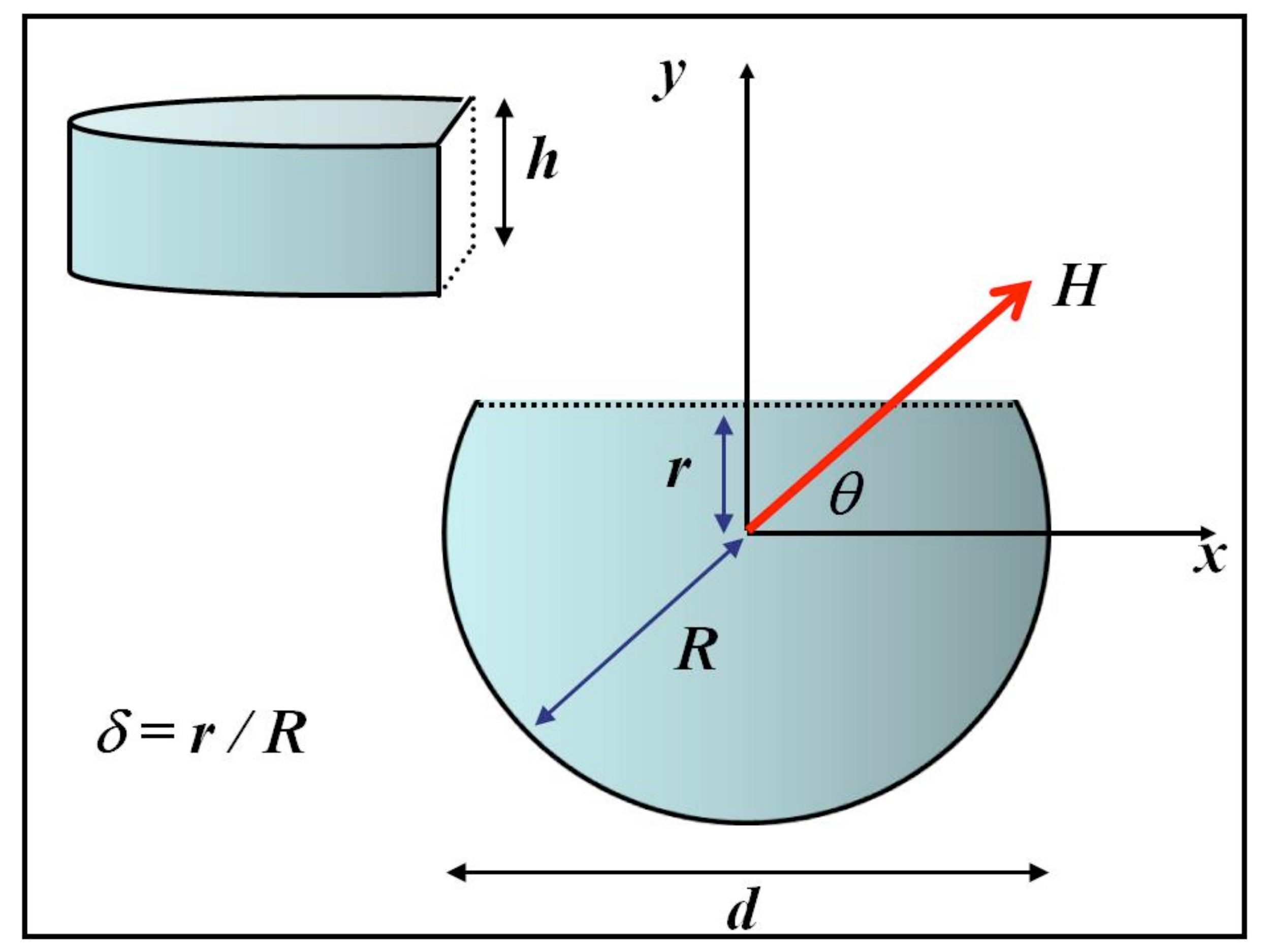}
\end{center}
\caption{(Color online) Geometrical parameters of a nanodot. The white
surface represents the section that has been cut.}
\end{figure}

\section{Results and Discussion}

\subsection{Magnetic field applied parallel to the x axis}

Our main concern in this work is to investigate the role of the asymmetric
shape of the disk in the coercivity and reversal mechanisms of the
magnetization. The hysteresis curves for different diameters and $\delta $,
with $h=20$ nm, are illustrated in Fig. 2, where we see for
all diameters an almost square loop with coercivities strongly dependent on
the geometry. For all diameters and $\delta =1.0$ the dots exhibit a small
coercivity, which increases while decreasing $\delta$.

\begin{figure}[h]
\begin{center}
\includegraphics[width=7cm]{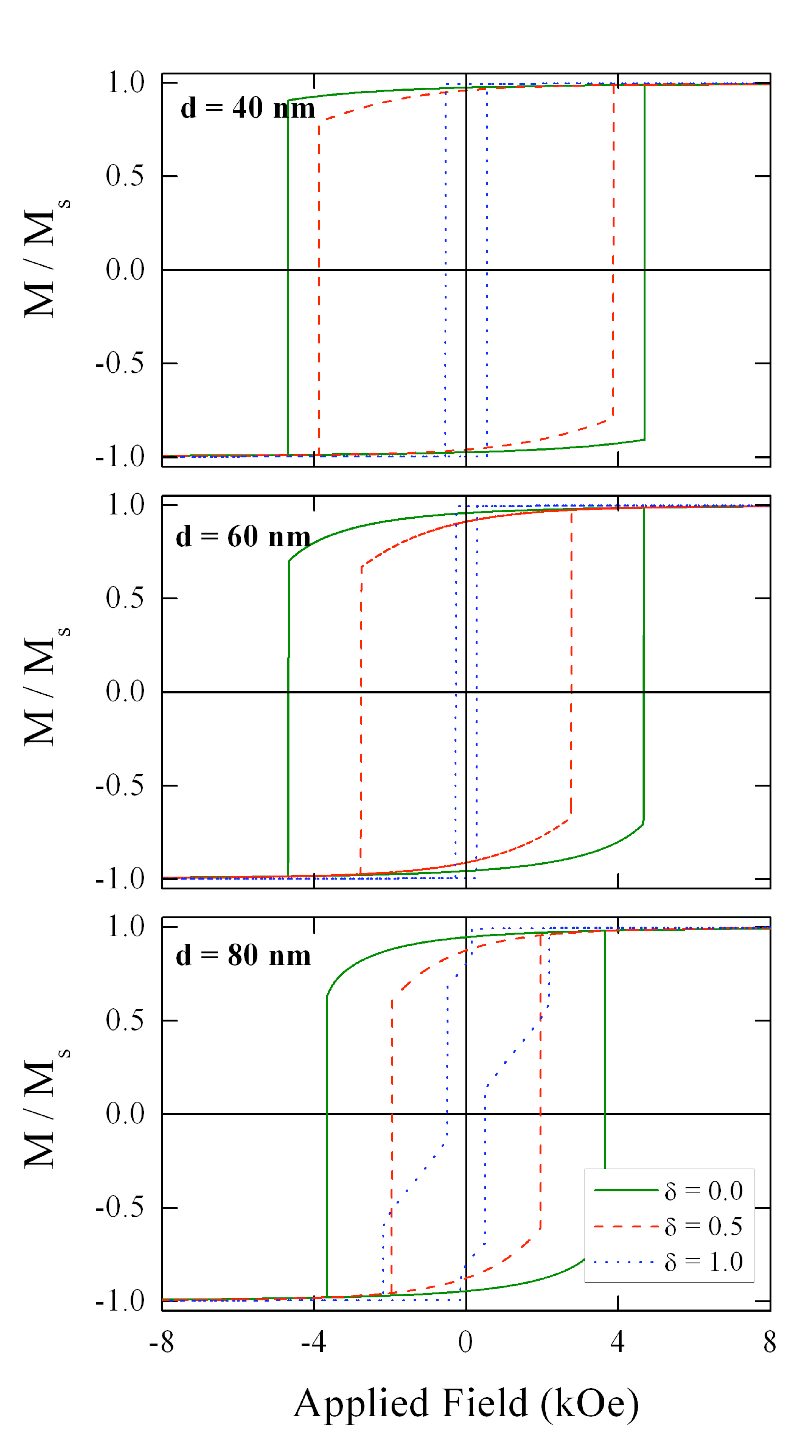}
\end{center}
\caption{(Color online) Hysteresis loops for asymmetric dots with height $%
h=20$ nm for different diameters and $\protect\delta$ values.}
\end{figure}

In Figs. 3 and 4 we illustrate the general behavior of a dot for different
values of $h$ and $\delta$. In these figures we observe a non-monotonic
behavior of coercivity and remanence which is the result of a competition
between exchange, local dipolar interactions and geometry in the region where the cut is
made. Once we have a small asymmetry ($\delta = 0.9$) in some cases a
decrease of the coercivity and remanence is seen. However, for further
decreases of $\delta$, remanence and coercivity are almost constant or grow
continuously.

\begin{figure}[h]
\begin{center}
\includegraphics[width=7cm]{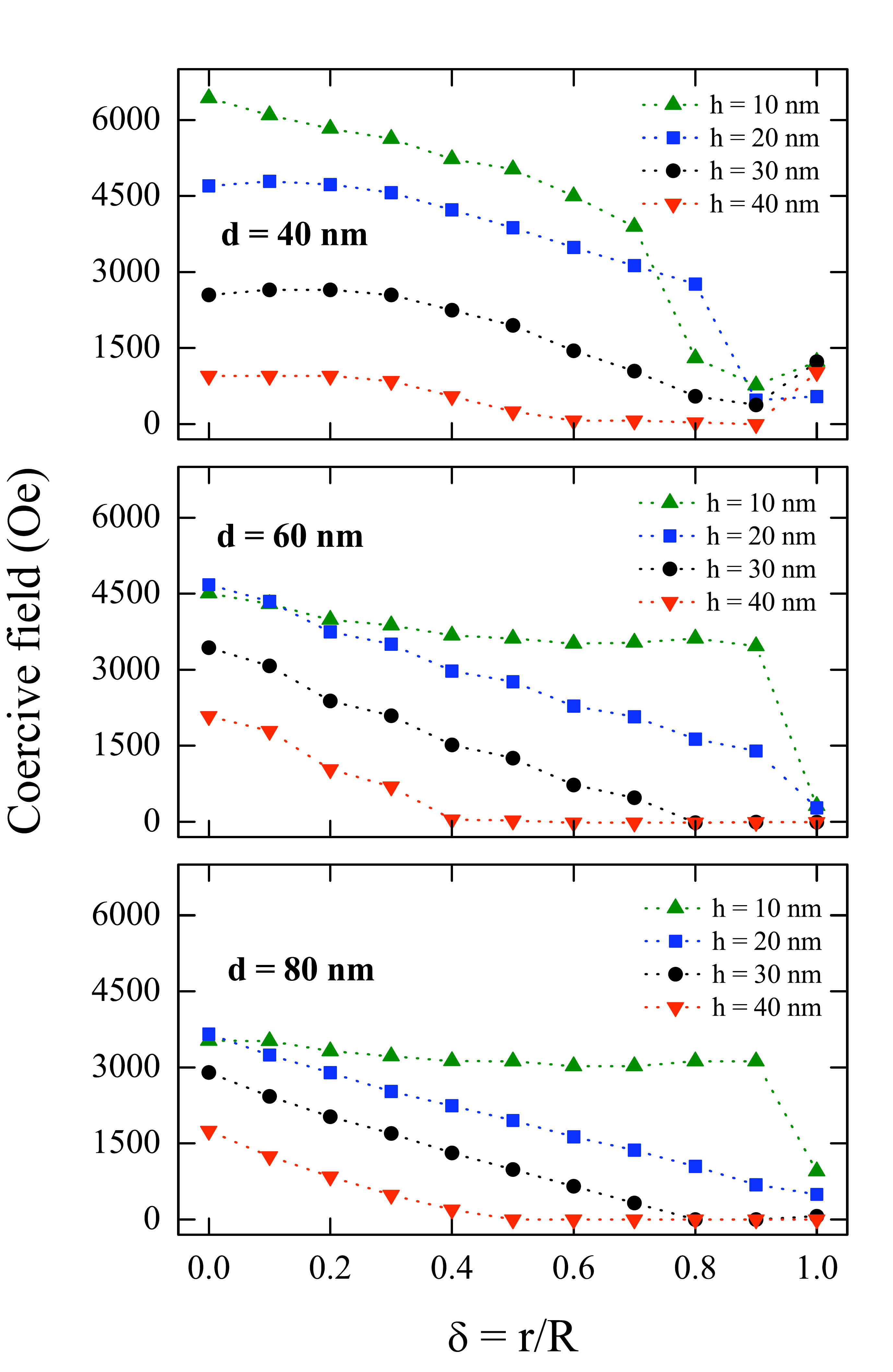}
\end{center}
\caption{(Color online) Coercivity of asymmetric dots for different height $%
h $, diameter $d$ and $\protect\delta$ values.}
\end{figure}

\begin{figure}[h]
\begin{center}
\includegraphics[width=7cm]{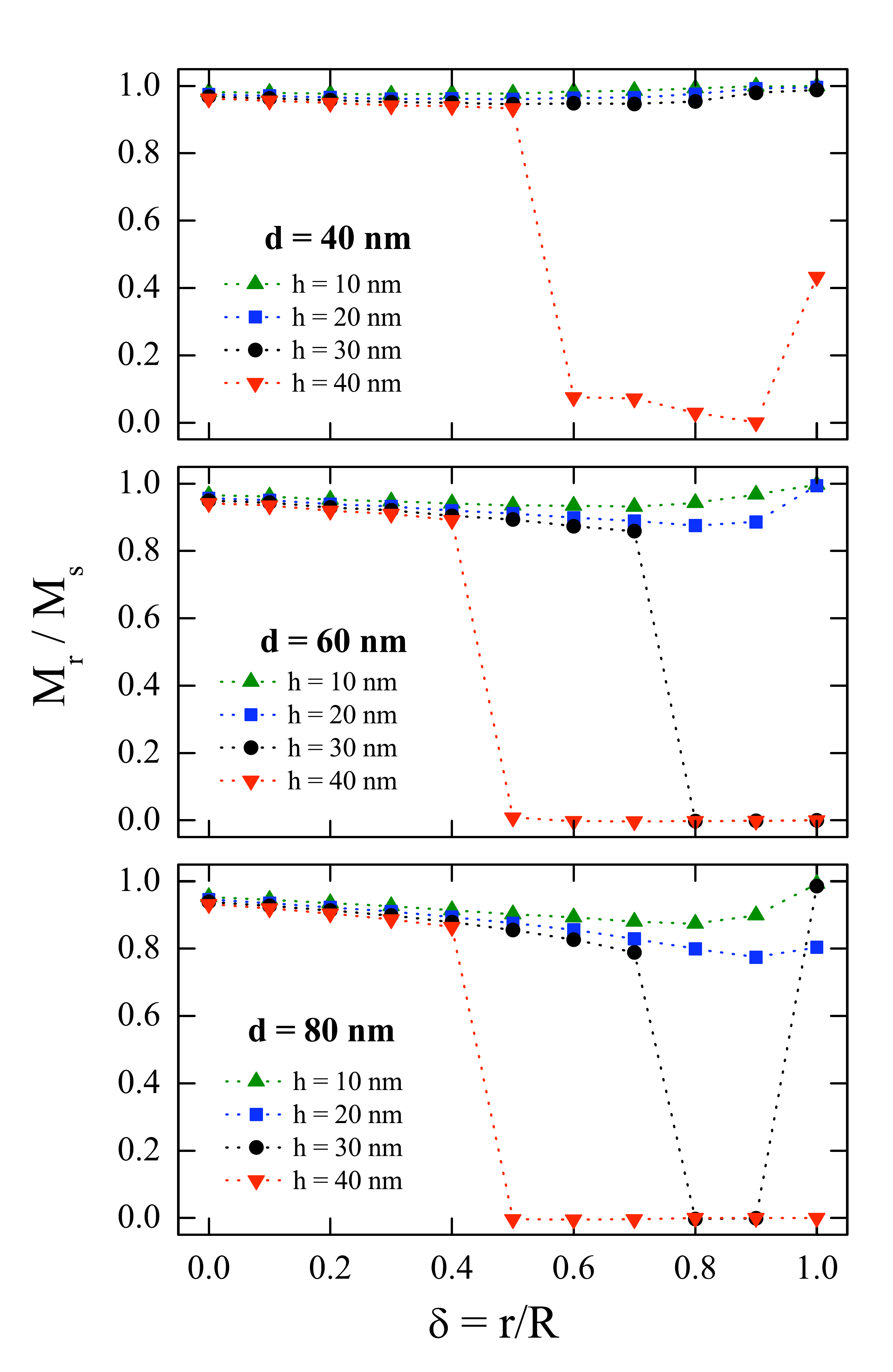}
\end{center}
\caption{(Color online) Remanence of asymmetric magnetic dots as a function
of $\protect\delta $.}
\end{figure}

In order to understand this behavior we look at snapshots of the hysteresis
and observe that the magnetization reversal occurs by a C formation followed
by vortex nucleation and propagation. This behavior is illustrated in Fig. 5(a)
for $d = 60$ nm, $h=20$ nm and each of the three values of $\delta$ used in Fig.
2. Looking at these snapshots we can conclude that a vortex nucleates during
the reversal for any value of $\delta$. In symmetric dots, square loops are a sign of coherent reversal, and the appearance of a neck indicates that the reversal is driven by a vortex nucleation and propagation\cite{MAL+10}. However for asymmetric dots reversal by vortex nucleation may lead to a square loop.

\begin{figure}[h]
\begin{center}
\includegraphics[width=7cm]{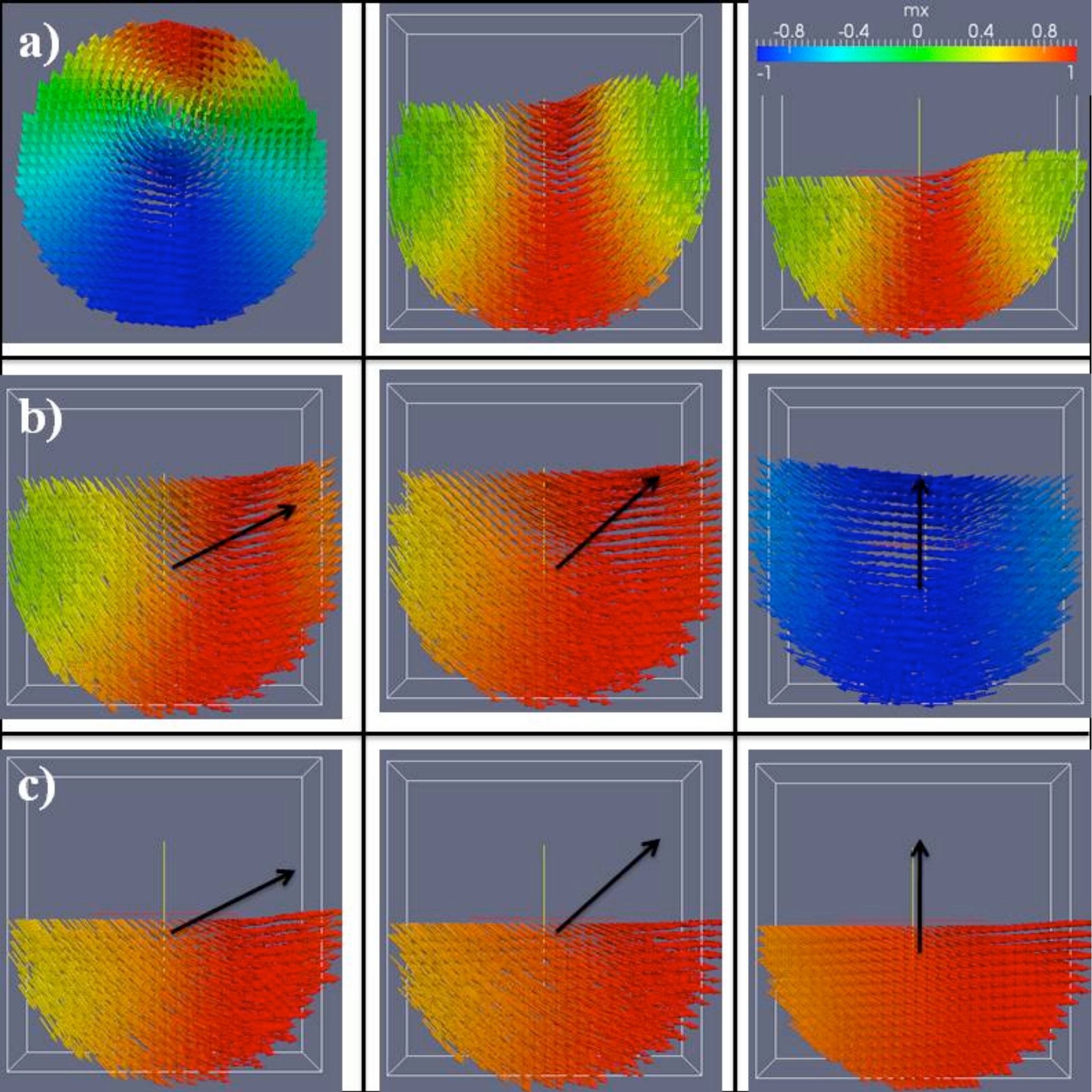}
\end{center}
\caption{Snapshots of the magnetization of the dot during the reversal for
a) $\protect\delta $=1.0, $\protect\delta $ = 0.5 and\ $\protect\delta $ =
0.0 when $\protect\theta $= 0, b) \ $\protect%
\theta $=30, 60 and 90 when $\protect\delta $ = 0.5, and c) \ $\protect\theta 
$=30, 60 and 90 when $\protect\delta $ = 0.0. The arrows illustrate the direction of the magnetic
moments. }
\end{figure}

Once a cut is made on the dot, a magnetic pole is formed on the new surface.
Due to the Pole Avoidance Principle\cite{Aharoni96}, a C state nucleates in
order to avoid the formation of a magnetic pole, followed by the nucleation of a vortex.
However, when the asymmetry increases by cutting larger sections of the dot,
the local effects described above compete with the new geometry, that is,
the lack of circular symmetry, making vortex formation more difficult. Then,
in this last case, C formation and vortex nucleation become more difficult
compared with the situation when a small cut is made or a symmetric dot is
considered, and an increase in coercivity and remanence appears.

\subsection{Angular dependence of the coercivity}

We also investigated the angular dependence of the magnetization, applying a
magnetic field along a direction defined by $\theta$, which is the angle
between the applied field direction and the $x$ axis, as illustrated in
Figure 1. Our results for $d$ = 60 nm and $h$ = 20 nm are depicted in Figs.
6 and 7. While symmetric dots show no angular dependence, as expected,
asymmetric dots exhibit a strong angular dependence. By increasing $\theta$
the coercivity decreases drastically until zero and a non
hysteretic behavior is observed for $\theta = 90^{o}$. This behavior allows us to conclude that shape anisotropy of an asymmetric dot may induce a hard axis of magnetization when $\theta = 90^{o}$. 

Besides, the magnetization inside an asymmetric dot is oriented parallel to the $x$ axis (easy axis), due to the shape anisotropy. When the applied field is reduced to zero, at remanence, each dot presents its magnetization along the axis, but it is measured at an angle $\theta$ with respect to the easy axis. Then, one can approximate the remanence of an asymmetric dot by $M_{R}(\theta)$=$M_{R}cos^{2}(\theta)$, with $M_{R}=M_{R}(\theta=0)$ the remanence measured at $\theta$ = 0. This behavior is an indication that asymmetric dots follow the behavior of uniaxial systems. Finally it is important to note that results for the remanence for $\delta$ = 0.0 and 0.5 almost coincide.

\begin{figure}[h]
\begin{center}
\includegraphics[width=7cm]{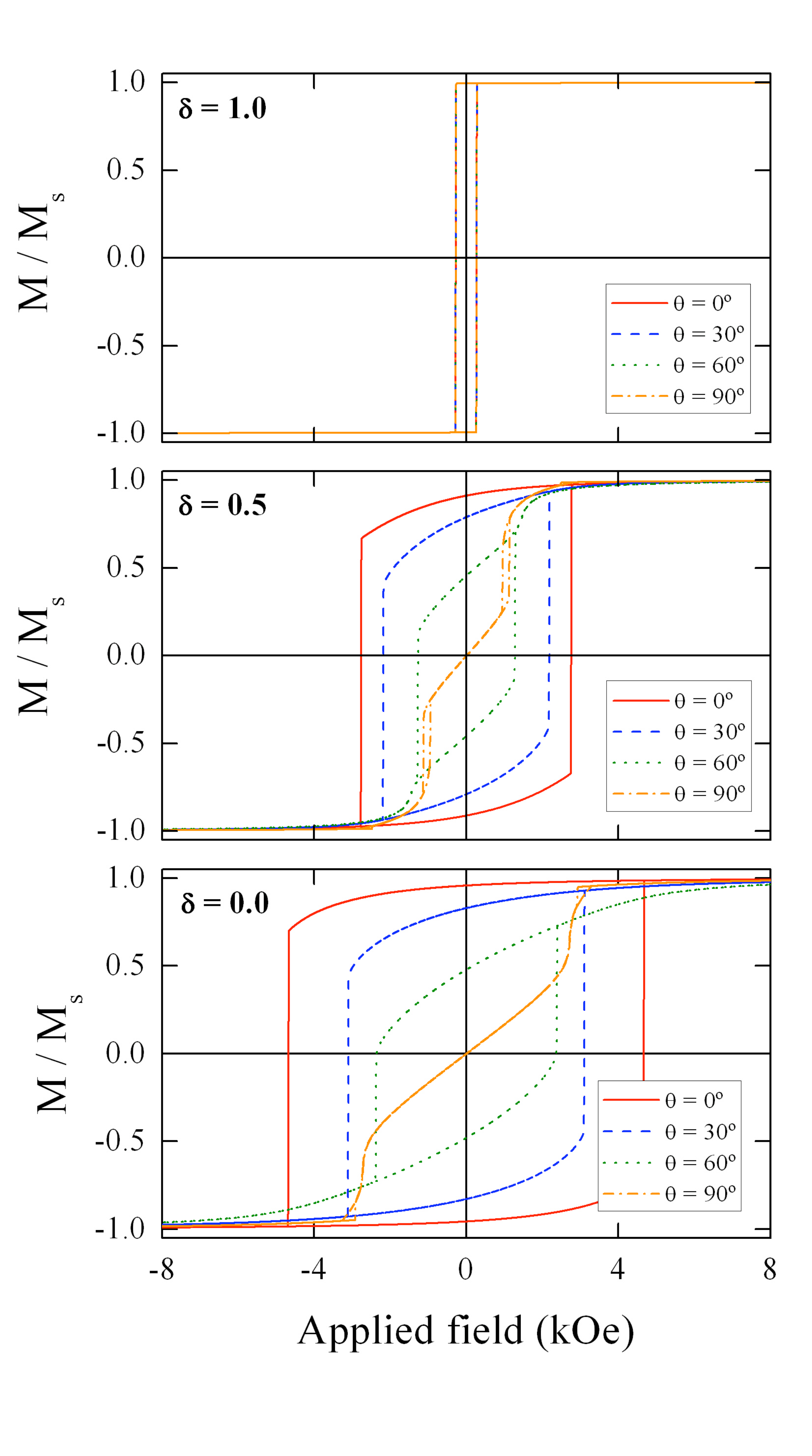}
\end{center}
\caption{(Color online) Hysteresis loops for asymmetric dots with diameter $%
d $ = 60 nm and height $h=20$ nm for different angles and $\protect\delta$
values.}
\end{figure}

\begin{figure}[h]
\begin{center}
\includegraphics[width=7cm]{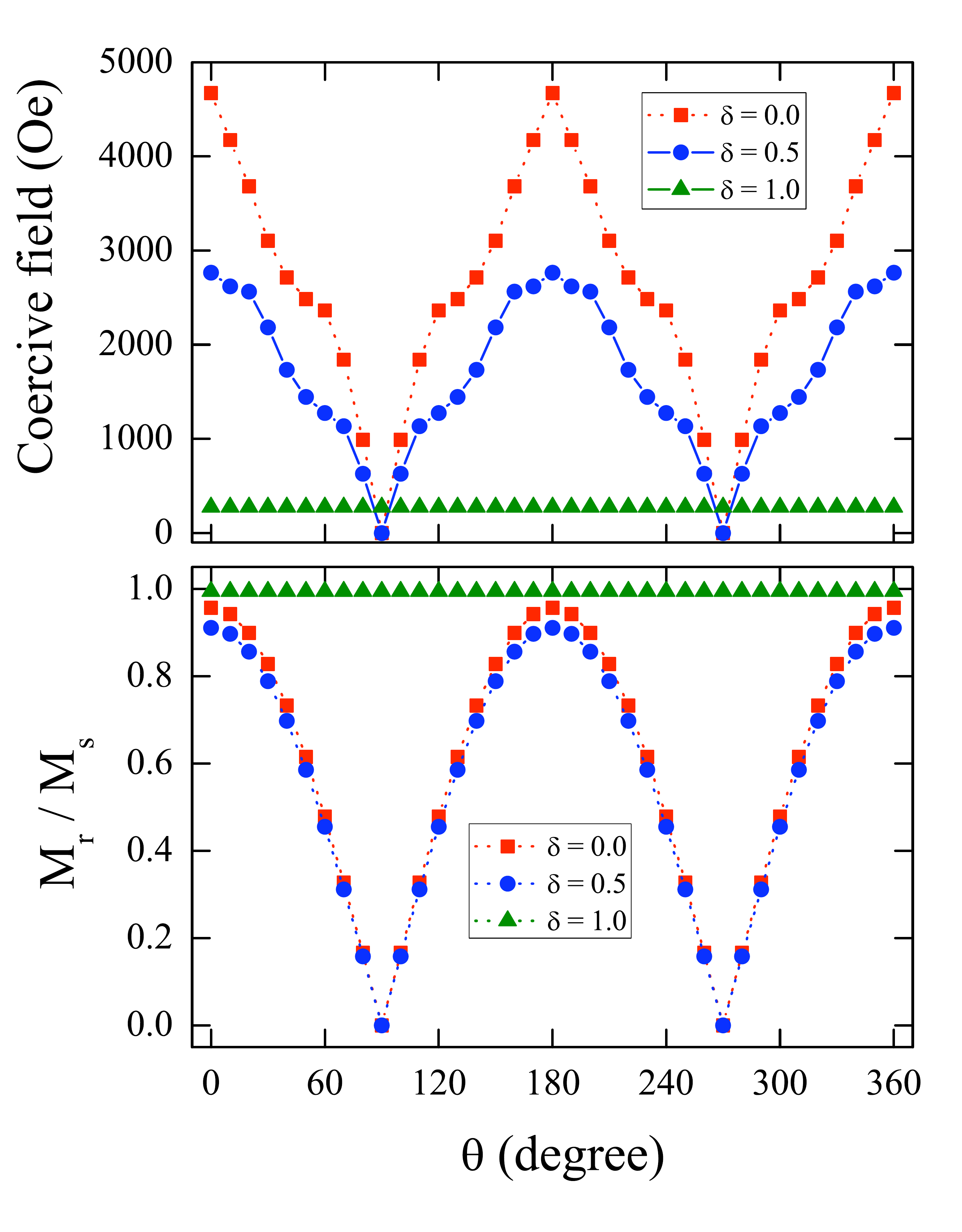}
\end{center}
\caption{(Color online) Coercivity and remanence of asymmetric dots with
diameter $d$ = 60 nm and height $h=20$ nm for different angles and $\protect%
\delta$ values.}
\end{figure}

To understand this behavior, snapshots of the reversal of the magnetization
are depicted in Fig. 5(b) and 5(c). It is seen that for $\delta=0.5$ the
reversal occurs by means of the nucleation and propagation of a vortex. For 0%
$^{o} \leq \theta < 30^{o}$, the vortex nucleates at the center of the
horizontal surface, and propagates along the $y$ direction until it is
annihilated at the opposite end of the dot. For 30$^{o} \leq \theta \leq
60^{o}$, the vortex nucleates at the center of the horizontal surface and
propagates along the $x$ direction until it is annihilated at one end of the
horizontal surface. For 60$^{o} < \theta \leq 90^{o}$, a coherent reversal
occurs.

For $\delta = 0.0$ the reversal of the magnetization occurs through the
nucleation of a C state and further nucleation and propagation of a vortex.
For 0$^{o} \leq \theta \leq 60^{o}$, two regions of different magnetic
orientation appear. These regions start growing until the full inversion of
the magnetization occurs. For 60$^{o} < \theta \leq 90^{o}$, a coherent
reversal occurs. The non monotonic behavior of the coercivity in Fig. 7 can be explained by the existence of different reversal
modes as a function of $\theta $.

\section{Conclusions}

The results presented above show that the existence of asymmetry strongly modifies the magnetic behavior of a dot. In this case the
coercivity and remanence are drastically modified as a function of $\delta$
when a magnetic field is applied parallel to the x axis. Also we have extended our results to the
case of an angular depedence of the coercivity and remanence, where a
transition from vortex-mode to coherent-mode rotation has been observed. In
this way asymmetry can be useful for tailoring specific magnetic
characteristics of these systems. However, experimental work remains to be
done in order to observe this transition.

\section*{Acknowledgments}

Support from FONDECYT under projects 11070010 and 1080300; the Millennium
Science Nucleus \textquotedblleft Basic and Applied
Magnetism\textquotedblright\ P06-022-F; and Financiamiento Basal para
Centros Cient\'{\i}ficos y Tecnol\'{o}gicos de Excelencia under project FB0807 is gratefully acknowledged.

\end{document}